# Understanding genomic alterations in cancer genomes using an integrative network approach

Edwin Wang[1, 2]

1. Lab of Bioinformatics and Systems Biology, National Research Council Canada, Montreal, Canada
2. McGill University Center for Bioinformatics, Montreal, Canada

**Abstract**

In recent years, cancer genome sequencing and other high-throughput studies of cancer genomes have generated many notable discoveries. In this review, Novel genomic alteration mechanisms, such as chromothripsis (chromosomal crisis) and kataegis (mutation storms), and their implications for cancer are discussed. Genomic alterations spur cancer genome evolution. Thus, the relationship between cancer clonal evolution and cancer stems cells is commented. The key question in cancer biology concerns how these genomic alterations support cancer development and metastasis in the context of biological functioning. Thus far, efforts such as pathway analysis have improved the understanding of the functional contributions of genetic mutations and DNA copy number variations to cancer development, progression and metastasis. However, the known pathways correspond to a small fraction, plausibly 5-10%, of somatic mutations and genes with an altered copy number. To develop a comprehensive understanding of the function of these genomic alterations in cancer, an integrative network framework is proposed and discussed. Finally, the challenges and the directions of studying cancer omic data using an integrative network approach are commented.

Keywords:

Chromothripsis

Kataegis

Cancer genome evolution

Network

Systems biology

Cancer genome sequencing

# 1. Introduction

Because of the ongoing development of new, fast, and inexpensive DNA sequencing technologies, the cost of genome sequencing technology has rapidly declined. Eventually, genome sequencing technology may allow doctors to decoding the entire genetic code of a patient disease sample in a clinical setting. Cancer is rooted in genetic and epigenetic alterations, which are either inherited or acquired (i.e., mutated or methylated) during our lives. Cancer genomes contain two classes of mutations: driver mutations, which are positively selected because they are essential for tumor growth and development, and passenger mutations, which are not subject to selection because they do not offer a growth advantage. Positive selection indicative of driver mutations is evidenced by a higher probability than by chance of amino acid-changing nonsynonymous mutations to synonymous mutations that do not involve amino acid changes.

However, we are still no closer to uncovering the molecular underpinnings of the disease. If we could catalog cancer driver-mutating genes, we could link these genes to biological pathways, biological processes and cellular networks. Cancer genome sequencing and omic studies will improve the understanding of the genetic/epigenetic underpinnings of major types and subtypes of human cancer, thus generating new insights concerning how to address and to manage cancer in clinics. These studies will eventually enable more personalized treatments that will improve patient care.

Significant effort has focused on compressive omic analysis, including genome sequencing, SNPs profiling, gene expression, epigenetic changes and proteomic profiling of major tumor types. These efforts represent a decade-long program that has aimed to generate high-quality omic data on more than 25,000 tumors for up to 50 types of cancer of clinical and societal importance. Three major organizations, the Wellcome Trust Sanger Institute in the United Kingdom, the Cancer Genome Atlas (TCGA) in the United States and the International Cancer Genome Consortium (ICGC) [1] in Canada, have advanced this program. Furthermore, several countries are working on this project, and the ICGC is collecting, merging and managing all of the data. In fact, the sequencing of 25,000 tumors is being performed around the world in 40 projects focused on the bladder, the blood, the bone, the brain, the breast, the cervix, the colon, the head and the neck, the kidney, the liver, the lung, the oral cavity, the ovary, the pancreas, the prostate, the rectum, the skin, the soft tissues, the stomach, and the uterus. The ICGC has collected omic data on more than 3,500 tumors.

Among these efforts, St. Jude Children's Research Hospital took an initiative to make cancer genome information freely accessible to the global scientific community at [www.ebi.ac.uk/ega/organisations/EGAO00000000046](www.ebi.ac.uk/ega/organisations/EGAO00000000046). [2] This site mainly includes the genome sequences (entire genome sequencing with up to a 30-fold coverage) for matched sets of normal and tumor tissue samples from pediatric cancer patients. It is expected to sequence more than 1,200 genomes by the end of 2012. It represents the world's greatest effort and investment to understand the genetic origins of childhood cancers and is the first major privately funded human genome sequencing project to share its data as soon as they become available. Table 1 summarizes the major completed genome sequencing efforts (i.e., at least 20 genomes have been sequenced in each study), most of which were published in 2012. In terms of cancer genome sequencing, 2012 signals the start of a significant increase in the amount of cancer genome sequencing data.

These efforts have generated several notable insights, including new genomic alteration mechanisms, the tracing and deciphering of cancer genome evolution and the dissection of tumor subclones.

2. **Cancer genome sequencing uncovered unexpected genomic alteration mechanisms**

*2.1. Chromothripsis*

Traditionally, cancer has been believed to develop gradually in a steady and stepwise progression. It often takes years to accumulate the multiple mutations to facilitate cancer's aggressive growth. In fact, most cancers undergo phases of abnormal tissue growth before developing into malignant tumors. However, recent analyses of tumor genomes revealed different patterns of genomic alteration mechanisms. One mechanism, "chromothripsis" [3], is a process by which a genome shatters into hundreds of fragments for one chromosome or several chromosomes in a single cellular catastrophe, and the DNA repair machinery then recombines the fragments in a highly erroneous order. This process generates gene amplifications/deletions on a massive scale. This mechanism suggests that a chromothripsis (also called a chromosomal crisis) can generate multiple oncogenes or tumor suppressors in a single event. Chromosomal crises have been observed across all common cancers at an incidence rate of 1/40 but are more common in bone cancers, with an incidence rate of 20% [3]. Chromothripses have been observed in other studies, although at a similarly low incidence rate (2%-4%) [4, 5].

It has been suggested that P53 mutation may trigger a chromothripsis. However, a P53 mutation is very common in cancer. An intriguing question concerns whether a chromosomal crisis occurs often or occasionally in the human body's cells. Notably, some cells that undergo a chromosomal crisis are able not only to survive but also to emerge with a genomic landscape that grants a selective advantage to the clone, thus promoting cancer development. Because a chromosomal crisis induces many dramatic genomic changes, we expect that a fraction of the genetic alterations are harmful to cells' survival. Should this expectation be accurate, many cells undergoing a chromosomal crisis could die, and we would be unaware of the occurrence. Another possibility is that tumor cells with many dramatic genetic alterations induced by a chromosomal crisis are so weak that they can be eliminated by tumor microenvironments. These factors may explain the extremely low rate of chromothripsis in tumors. Comparing the microenvironments of the tumors that have or do not have chromothripsis might offer insight regarding this point. The survival outcomes of the patients with chromothripsis-derived and patients with non-chromothripsis-derived tumors need to be reviewed.

*2.2. Kataegis*

Chromothripsis refers to a new oncogenic mechanism that operates on a global level; that is, it occurs in one chromosome or in several chromosomes. Another new mutation mechanism has been found to operate locally, generating large numbers of mutations (or hotspots of hypermutations) in small regions of the genome; this process has been termed 'kataegis' (from the Greek word for thunderstorm) [6]. Compared with chromothripsis, which occurs at a rate of 1/40 in cancer genomes, kataegis is remarkably common, occurring at a rate of 13/21 in breast cancer genomes. Kataegis most likely causes mutations to occur at one time rather than accumulating in a step-wise fashion. Kataegis is observed when a large number of substitution mutations occur simultaneously and extremely close to one another in a region of the genome [6]. Therefore, kataegis can be identified as distinguishing mutational patterns that often co-occur with large-scale rearrangements. Kataegis has been described as most likely occurring at one time; however, certain regions of the genome may mutate repeatedly during tumor evolution. If this possibility is proven, the affected regions are extended due to mutations between different subclones. Whether such extended mutation profiles are found in these genome regions requires further examination.

These mutation mechanisms indicate that internal factors, such as genome rearrangement, may contribute some types of genetic mutations and that some regions of cancer genomes are more likely to mutate than others. Several questions remain. The first question concerns (1) whether kataegis occurs once or repeatedly in different subclones. Based on the results, kataegis most likely occurs in different subclones across the tumor's evolution; however, this conclusion need to be confirmed. The remaining questions concern (2) whether the genomic alterations derived from kataegis or chromothripsis are enriched with cancer drivers; (3) whether kataegis or chromothripsis results from specifically defective internal cellular machinery or certain combinatory patterns of existing mutations or from other epigenetic alterations; (4) whether kataegis-derived cancer genomes are more aggressive than the tumors derived from other genomic alteration mechanisms; and (5) whether the DNA copy number change represents an important genetic alteration in cancer cells. A series of chromothripses might convert a cell into a cancer cell quickly. However, these cells may not mutate but rather undergo many copy number changes. Further investigations of these novel genomic alteration mechanisms may deepen the understanding of factors influencing these mutations/copy number variations and cancer development. Nevertheless, early mutations, copy number changes or alterations of chromosomal structures caused by other factors, such as smoking, sunlight, or stress, may be conducive to kataegis or chromothripsis.

**3. Genomic alterations and cancer genome sequencing**

*3.1. Deciphering cancer genomic alterations in an evolutionary context*

Tumors are widely known to be heterogeneous. Therefore, a tumor is often expected to comprise a set of clones of cancer cells. Understanding the mechanisms that cause cancer requires comprehensive insight into the processes underlying cancer genetics. To obtain this information, the Sanger group sequenced 21 breast cancer genomes, catalogued their mutations and attempted to trace the evolution of cancer mutation profiles between different subclones in the context of timing [7]. Using appropriate bioinformatic approaches, they were able to uncover the genetic mutation marks and the inscribed record of the latter's emergence over time, thus enabling the divergence of a cell in forming the various clones to be traced.

Cancer mutational processes evolve over a tumor's lifespan. Each time a gene is mutated in a cell, the mutation event encrypts a distinctive stamp inscribed in its genome. Such a distinctive stamp can be considered a mutation signature or a distinctive mark on the genome. From a historical perspective, an archaeological history of mutations results from the accumulations of marks upon one another. Therefore, collecting, cataloguing and sorting the orders of these marks allow the cancer's life history to be deciphered.

Cells in a single subclone possess an identical genetic mutation profile. The cells in late-occurring subclones add new mutations to the genetic mutation profiles of the early subclones. Only a mutation causes a clone to dramatically expand, after which the clone constitutes the tumor's dominant population and the tumor grows large enough to be diagnosed. During this process, the history of the genetic mutations and the emergence and the expansion of the new clone is imprinted upon the cancer genome. Therefore, these genetic mutation profiles can be used to build an evolutionary tree of subclones. This analysis allows the tracing of the occurrence of a new mutation profile, the biological processes involved in each mutation profile or subclone, and the proportion of each subclone present in the tumors. The study showed that as the mutations, or genetic marks, accumulate in a cell, the corresponding changes caused the cancer to diverge into different clones [7]. The emergence of different subclones occurred in all of the examined cancer genomes. This analysis suggests that when a new subclone has grown sufficiently to constitute the dominant population of cancer cells (i.e., more than 50% of tumor cells), the tumor becomes clinically detectable. The genetic mutation profiles of subclones and dominant subclones differ markedly, implying that the dominant subclone may require a significant amount of time to develop.

   This study has several implications. First, it confirms that a tumor contains many subclones; thus, the heterogeneity can be sampled by sequencing, as shown in Gerlinger et al. [7]. Second, the study suggests that a cancer patient may benefit from the use of combinatory drugs for different subclones. One perspective is that treatment may proceed by targeting the early clones because their mutation profiles are identical to those of the late subclones [8]. However, this approach might not work well. The late subclones, particularly the dominant subclones, acquire new traits for cell proliferation and survival. These new traits could grant the late subclones "new" survival capabilities, thus implying that the 'drug targets' for the early subclones are unsuitable for the late subclones. However, subclones [9] can successfully be dissected from tumors, and the RNA-seq or sequencing analysis of these subclones can be performed [10]. Profiling these subclones helps to identify drug targets for each subclone. Third, each cancer appears to be generated by a distinct combination of mutational processes. Fourth, the study confirms the prediction that transforming a normal cell to a cancer cell is a lengthy process that has been shown to last 10 years [11]. This fact supports the search for mutation signatures or molecular biomarkers for early diagnosis and the possibility of personalized early preventions to potential cancer patients.

*3.2. Which subclone contains the cancer stem cell?*

Although many implications have been found from deciphering cancer genomes, a key question remains: can we identify or infer which subclone contains the cancer stem cell?

The cancer stem cell hypothesis suggests that tumors are organized into aberrant cell hierarchies. In such a hierarchy, a cancer stem cell capable of replicating indefinitely is the parent cell that generates differentiated daughter cells with a limited capacity to proliferate. If proven for a tumor, this hypothesis suggests that cancer is curable if drugs are available to kill cancer stem cells. Recently, three studies have provided the first evidence that cancer stem cells exist [12]. These studies have demonstrated that cancer stem cells reproduce cancer cells when the tumor is extinguished by anticancer drugs, a result that applies to benign and malignant tumors.

These results suggest that cancer stems cells are inherently capable of propagating genetic or epigenetic alterations throughout a tumor and driving cancer evolution. The cancer stem cells constitute the stem of tumor cell production. During a cancer's evolution, a certain genetic mutation profile may define cancer stem cell subclones. The parental subclones of the cancer stem cell subclone are non-stem cells, thus indicating that genetic/epigenetic changes could convert a non-stem cell into a cancer stem cell. Such a stem cell produces many daughter cells that may further evolve into diversified cancer cells. If this process occurs, then the dominant subclones of the tumors would not be cancer stem cells but would emerge from cancer stem cells.

Thus, many questions remain. First, should every tumor contain a subset of cancer stem cells? All tumors may not contain cancer stem cells, for example, if a chromothripsis occurs to generate enough fitting combinations of mutations to generate a new subclone capable of becoming dominant in a tumor. However, generally, sustaining the cancer's evolution requires generating enough cancer cells using cancer stem cells. In this case, a cancer stem cell must evolve and be defined by a set of mutations. Second, if a cancer stem cell must exist in a tumor, what are the characteristics of its genetic mutation profiles? The evolution of cancer genomes implies that a cancer stem cell subclone should be an early subclone in the history of the tumor's development. We assume that the cancer stem cell subclone generates new cells but at a limited and slow rate because of the microenvironment's unfavorable conditions. These daughter cells are further mutated and generate new subclones.

This hypothesis is supported by the discoveries of Driessens *et al*. [12-14] and Chen *et al*. [13] that show that the cellular organization of early (precancerous) skin and intestinal tumors are composed of both stem cells and non-stem cells. Furthermore, Driessens *et al*. observed that progression to cancer in benign skin tumors is associated with an increased cancer stem cell population. They found that every cancer stem cell within a tumor is equally probable of clonally expanding or dying, possibly even in the absence of new mutations, thus suggesting that a cancer stem cell subclone most likely occurs in an early stage of cancer development (Fig 1).

The most important finding from the above discussion is that subclonal diversity in cancer genomes may be dissected. To treat cancer more efficiently requires the dissection of not only the dominant subclones but also the minor subclones, which are among the top subclones in the hierarchy and most likely cancer stem cells. The key challenges are determining the functions of the mutations in these subclones and linking these mutations to drug targets for each subclone and, subsequently, identifying mutational signatures as diagnostic tools that predict the early development of cancer.

## 4. The functional understanding of genomic alterations in cancer genomes at a systems level

As discussed above, in recent years, sequencing efforts have generated many genomic alterations of cancer. Although some important insights, such as new genomic alteration mechanisms, cancer genome evolution, and clonal dissection, have been gleaned from the corresponding data, a lack of frameworks and tools nevertheless persists with respect to how genomic alterations drive cancer evolution, clonal expansion, and cancer development and metastasis. Understanding these processes requires investigating the function of these genomic alterations. However, currently, the greatest challenge is understanding the functions of driver mutations in cancer genomes. On average, several hundred driver mutations have been catalogued from each tumor, but interpreting these mutations is inefficient, thus limiting the understanding of why and how these mutation genes work. Most current cancer genome sequencing work has mapped the mutations onto biological pathways, including signaling pathways.

A biological pathway contains a set of linear relationships between genes or proteins and is an important model to describe biological relationships. From an abstract perspective, a pathway is a model that represents humans' efforts to explore, describe, and organize biological relationships in cells [15]. Such efforts allow us to further understand biological systems and to predict cell behaviors. However, less than 5-10% of mutations can be explained by these 'core pathways' in which a set of genes is relatively highly mutated [16, 17]. Several problems are associated with this approach. First, pathway gene members are fuzzy and poorly defined. For example, if different databases such as KEGG or BioCarta are queried for 'EGFR signaling pathway,' each database will provide a different gene list. Second, only a small set of mutation genes (less than 5-10%) can be mapped onto the 'core pathways,' thus, most mutation genes are unmapped. Third, for each study, only a few 'core pathways' such as RAS-PI3K were found repeatedly. These pathways are well known in cancer studies, and drugs have been developed for these pathways or key genes. In this respect, the pathway approach does not provide new insights but captures the existing knowledge in the field.

To comprehensively understand the drivers of cancer genomes, computational methods should be developed because of the large number of driver mutation genes in each tumor. The development of effective computational methods first requires understanding what subjects biology addresses. Over many years, biologists have explored several types of relationships among genes, proteins, RNAs and other molecules. The molecular mechanisms of biological processes or diseases have also been studied. A molecular mechanism often discloses an event of A regulating B, B inducing or inhibiting C and generating a phenotype or disease. Each mechanism comprises a set of relationships between molecules such as gene regulation, protein interaction, activation, genetic interaction and inhibitory action. In this context, biology is a science that describes relationships [15].

Traditionally, biologists use pathways to describe the relationships between a limited number of genes. As shown above, numerous driver mutations have been found in each tumor, but the pathway approach explains only 5-10% of the genetic mutations in the cancer genome. It is expected that biologists must consider thousands of biological relationships with respect to these genes in one tumor, for which task the traditional methods of describing biological relationships are inadequate. Furthermore, as the cross-talk between pathways has been documented extensively in recent years, scientists have come to recognize that a network model may provide a solution. A network model describes the interactions and relationships between genes or proteins in a nonlinear manner. The pathway and network approaches are conceptual models that reflect an avenue of exploring biology by humans. Both models project living cell molecules onto conceptual frameworks, but the molecular network model more closely approximates the processes that occur in cells.

*4.1. Basic concepts of molecular networks*

A molecular network can be represented using nodes and links. Nodes represent genes or proteins, whereas links represent the relationships between genes and proteins. Cancer omic data can be integrated into a molecular network to pose questions and obtain corresponding insights. However, different types of networks have distinct properties and thus can be used to address different questions.

There are four basic molecular networks: signaling, gene regulatory, metabolic and protein interaction networks. A network hub consists of the nodes that have the most links (top 5-10% of the links) within the network. However, a hub assumes a different biological meaning in different networks: hubs in gene regulatory networks represent global transcription factors, which assume a major role in responding to stimuli and coordinating the regulated genes. Consistent with this mechanism, the transcripts of the hub transcription factors often rapidly decay [18]. The rapidly decaying transcripts of global hub transcription factors might encode "switch" functions, which are used under different conditions and stimuli. Hubs in cellular signaling and metabolic networks help to organize information flows within the networks. Hubs in signaling networks assume a major role in integrating different signals and pathways [19]. However, hubs in protein interactions indicate that hub proteins are involved in considerably more biological processes/protein complexes than are other proteins [20].

Most targets of transcription factors in regulatory networks are "workers." They directly execute the tasks of biological processes. Most nodes in gene regulatory network have no regulatory role, whereas most nodes in cellular signaling networks have regulatory roles. In signaling networks, approximately all nodes are "regulators." Therefore, logical regulatory relationships are extensively encoded in signaling networks. In another view, nodes in signaling networks are information carriers. Similarly, nodes in metabolic networks are carriers of the metabolic flux. Nodes in protein interaction networks are members of protein complexes, or 'co-workers,' that interact and collaborate to complete specific tasks. [15]

Generally, signaling networks contain several logical codes of regulation, whereas protein interaction networks do not encode regulatory logics. Gene regulatory networks encode both regulatory logic and gene "workers." Gene regulatory networks help to identify key cancer regulators, the co-expression of genes, and the sets of "workers" involved in cancer processes, whereas signaling networks help to identify cancer causal genes and the regulatory logic in cancer processes [15]. Protein interaction networks provide a source for identifying co-workers in cancer progression or metastasis [15]. In recent years, notable discoveries have also addressed cancer metabolism. For example, hypoxia and isocitrate dehydrogenase 1 (*IDH1*) function in the citric acid cycle pathway, an energy-generating pathway. The *IDH1* mutations cause the accumulation of 2-hydroxyglutarate, a metabolite in the metabolic pathway, and thus cause cancer [21] [22, 23]. These studies generate interest in studying cancer in metabolic networks. Metabolic networks could help identify key enzymes or metabolites that could induce metabolic chaos and thus cause cancer.

When incorporating cancer omic data into these networks, the expression levels of major regulators (i.e., kinases) in signaling networks do not necessarily change dramatically during cancer progression and metastasis because signaling cascades are modulated through protein modifications, not gene regulation. Indeed, examining phosphoproteomic data in signaling networks is informative [24].

*4.2. An integrative network analysis of cancer omic data*

Prior to the advent of cancer genome sequencing, researchers pursued network analyses of cancer-related data. Additional details regarding this research are available in reviews [25, 26] and in the book *Cancer Systems Biology* [15]. For example, an integrative analysis of tumor gene expression profiles and protein interaction networks permits the identification of network modules as cancer biomarkers. A network module contains a set of closely interacted genes/proteins that collaborate to perform certain biological process. In a network module, the interactions between the module members are more frequent than the interactions between the module members with the protein/nodes outside the module. The resulting network module markers are more reproducible than traditionally obtained biomarkers that are identified without using protein interaction networks [27, 28].

Another study found that the protein network modules of gene co-expression are more reliable predictors than the biomarkers that are identified without considering network modules [29]. Exhaustively describing human molecular networks is a long-term task. Therefore, although the network module approach has yielded promising results, the accuracy of its predictions is nevertheless low (less than 80%). However, excellent results (i.e., over 90% prediction accuracy when tested in multiple independent patient cohorts) have been obtained with biomarkers through cancer-hallmark-based functional modules that extend beyond the network module [30].

A functional module biomarker is linked to a network module of cancer driver-mutating genes on the protein interaction network. Notably, as long as the genes, which perform the same function of the module and act as the interacting partners of proteins in the cancer-driver network module, are modulated, they show predictive power (Figure 2) [30]. Most importantly, the members of the cancer-driver network module are fixed, but the modules' modulated interacting partners change with patient cohorts [30]. These results explain why many potential biomarker sets exhibit limited overlap, a well-known phenomenon in the cancer biomarker field.

The first study that attempted to interpret the drivers behind cancer-mutating genes in networks was published in 2007. In this work, Cui et al. [31] first examined the traditionally discovered cancer-mutating genes, which were considered to be cancer drivers and which were mined using the COSMIC database, using a manually curated human signaling network. They identified 12 cancer signaling network modules in which cancer-mutating genes were densely connected. Such modules cannot be discovered using a pathway approach or by examining mutating genes individually. Cui et al. applied these modules to the drivers behind the mutating genes derived from the first genome-wide sequencing of cancer [32] and found that these mutations became meaningful if assigned to 2 or 3 signaling network modules [31].

This result resolved the controversial arguments concerning whether tumor genome sequencing should continue [33] because the first study of tumor genome sequencing showed that cancer mutations are highly complex: driver-mutating genes are rare, even in the tumors of sample origin, such as breast cancer [34]. In addition, these genes are not organized into pathways. Subsequently, some scientists argued that cancer genome sequencing is time- and resource-consuming and provides unclear results [33]. Cui et al.'s network analysis provides evidence that cancer driver-mutating genes are organized into network modules to assume roles. Lincoln Stein and colleagues [35] used an extended signaling network to examine cancer driver-mutating genes derived from the genome sequencing of two glioblastoma multiforme data sets [17, 36] and confirmed that cancer driver-mutating genes are organized into network modules.

Given that current networks are incomplete, Ciriello et al. [37] used correlation analysis and statistical approaches to develop a method for identifying cancer gene modules, specifically, Mutual Exclusivity Modules in cancer (MEMo). MEMo considers the frequency of cancer gene mutation across a set of tumor samples. Cancer mutation genes are most likely to participate in the same biological process, and cancer-mutating genes that are involved in the same biological process are mutually exclusive in a module. This approach may help detect cancer gene modules that cannot be found with existing networks. However, a limitation of the method is that the sample number significantly affects the results.

Akavia et al. [38] used a Bayesian network approach to develop a network-module-based method, CONEXIC (computational algorithm, copy number and expression in cancer) to examine DNA copy number alterations in tumor genomes. This approach integrates copy number alterations and gene expression profiles to identify potential cancer driver genes that have recurrent copy number changes in tumor samples. A limitation of this study is that the Bayesian network learning approach and other network learning approaches exhibit weak performance for eukaryotic genomes, such as those in yeast and humans [39].

## 5. Future directions

An integrative network analysis of cancer genome sequencing data is in its early phases. Several challenges and questions need to be addressed. First, in terms of their functional importance, the cancer driver-mutating genes need to be prioritized into categories such as cancer cell survival or metastasis. This task is urgent because each tumor contains several hundred cancer driver-mutating genes, but it is unclear which genes should be prioritized in research. New network methods can be developed to answer this question. Second, the network of cancer stems cells requires further study to understand which combinations of mutating genes or epigenetic/copy number changes maintain the characteristics of cancer stems cells. Third, the network dynamics of cancer evolution should be mapped, and the key events that lead to cancer initiation need to be understood. Fourth, the networks of cancer hallmarks should be modeled. This approach has been proposed, and its necessity has been detailed in the book *Cancer Systems Biology* [15]. Li et al. obtained excellent results in identifying cancer biomarkers by modeling functional modules of cancer hallmarks [30]. Fu et al. also showed that apoptotic network motif modeling based one of the cancer hallmarks inferred that the 26S proteasome may help predict cancer prognosis [40]. Fifth, the dynamics of the cancer gene network motifs need to be studied, and the regulatory principles and controls of the oncogenic signaling information flows should be understood. In-depth discussions are available in a recent review [26]. Finally, new network algorithms need to be developed to translate some cancer mutations and other genetic/epigenetic changes into drug targets. Inarguably, many other problems and activities may be solved through an integrative network approach. With additional cancer high-throughput studies in the near future, we believe that the integrative network approach will become increasingly useful to cancer research.

## Acknowledgements

This work is supported by Canadian Institute of Health Research and NRC's Genome and Health Initiative.

**Figure Legends**

**Fig. 1**. A hypothesis of a stem cell-based cancer genome evolution. The production of a cancer stem cell requires a long time (5-10 years) and a series of changes with respect to somatic mutations, copy number variations and epigenetic factors from a normal cell. The cancer stem cell generates cancer cells. Some mutations or copy number changes could lead one cancer cell derived from the cancer stem cell to acquire significant growth advantages. This cancer cell will propagate to dominate the tumor and eventually make the tumor clinically detectable.

**Fig. 2**. Stable cancer driver-mutating gene modules have predictive power in cancer prognosis. The content of this figure was developed from a cancer prognostic gene signature derived from Li et al. [30]. For example, this gene signature represents apoptosis, one of the cancer hallmarks. A driver-module is formed by collecting the genes (blue) that are the interacting neighbors of the signature genes (brown). By obtaining microarray gene expression profiles from a breast cancer cohort, we obtained modulated genes between recurred and non-recurred tumors. We sought a predictive gene signature by collecting the genes (green) that are not only the interacting neighbors of the driver-module genes (blue), but also the modulated genes sharing the same function with the original gene signature (i.e., apoptosis). We found that the predictive gene signature (green genes) has predictive power for prognosis. It is notable that another set of the predictive gene signature (green genes) can be obtained when changing patient cohorts. However, the new genes in green maintain their predictive power. These results explain the phenomenon that the gene sets of different cancer biomarkers for an identical cancer type rarely overlap. More detailed explanations are in Li et al. [30]. Protein interaction networks contain circles (nodes) and solid lines that represent genes and their interactions, respectively. Brown, blue and green nodes represent signature genes, breast cancer driver-mutating genes, and breast cancer modulated genes, respectively, between recurred and non-recurred tumors. Dotted lines represent the signature genes mapped to the genes in the network.

**Table 1. A list of major completed tumor genome sequencing projects.**

| Tumor type | Number of tumors sequenced | Year | Reference |
|---|---|---|---|
| Lung cancer | 31 | 2012 | [41] |
| Lung cancer | 53 | 2012 | [42] |
| Lung cancer | 183 | 2012 | [43] |
| Lung cancer | 178 | 2012 | [44] |
| Breast cancer | 825 | 2012 | [45] |
| Breast cancer | 125 | 2012 | [46] |
| Breast cancer | 77 | 2012 | [47] |
| Breast cancer | 21 | 2012 | [7] |
| Breast cancer | 100 | 2012 | [48] |
| Breast cancer | 65 | 2012 | [49] |
| Breast cancer | 11 | 2006 | [50] |
| Medulloblastoma | 37 | 2012 | [51] |
| Medulloblastoma | 92 | 2012 | [52] |
| Medulloblastoma | 125 | 2012 | [53] |
| Acute myeloid leukemia | 24 | 2012 | [54] |
| Multiple myeloma | 38 | 2011 | [55] |
| Chronic lymphocytic leukemia | 105 | 2012 | [56] |
| Melanoma | 121 | 2012 | [57] |
| Melanoma | 25 | 2012 | [58] |
| Prostate cancer | 112 | 2012 | [59] |
| Prostate cancer | 50 | 2012 | [60] |
| Prostate cancer | 23 | 2011 | [61] |
| Colon and rectal cancer | 224 | 2012 | [62] |
| Colon cancer | 11 | 2006 | [50] |
| Liver cancer | 27 | 2012 | [63] |

| Hepatocellular carcinoma | 24 | 2012 | [64] |
| --- | --- | --- | --- |
| Diffuse large B-cell lymphoma | 55 | 2012 | [65] |
| Pediatric glioblastoma | 48 | 2012 | [66] |
| Gastric cancer | 22 | 2011 | [67] |
| Ovarian carcinoma | 316 | 2011 | [68] |
| Head and neck carcinoma | 74 | 2011 | [69] |
| Head and neck carcinoma | 32 | 2011 | [70] |

Note: Projects that have sequenced more than 20 cancer genomes at an exome or an entire genome level have been selected. The majority of the work was published in 2012.

Fig 1

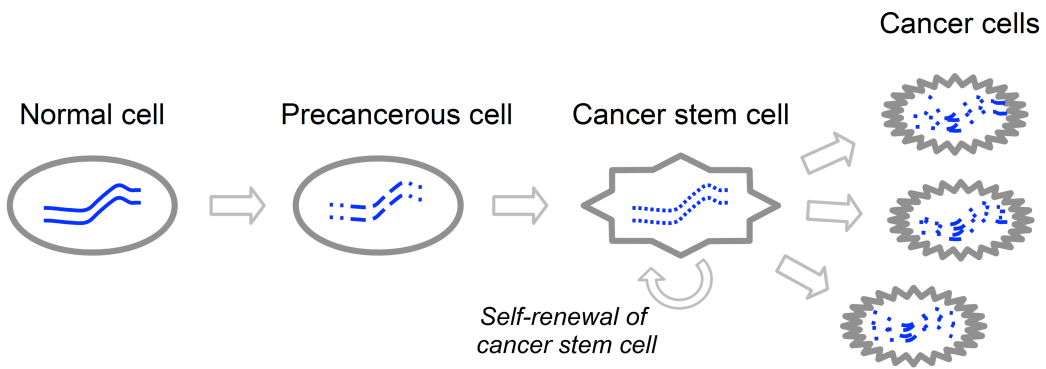

Fig 2

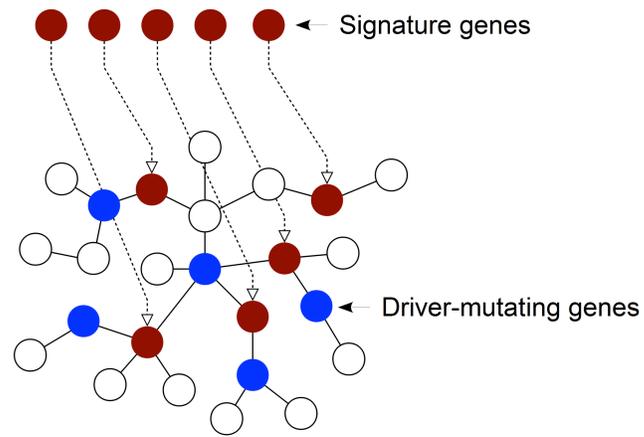
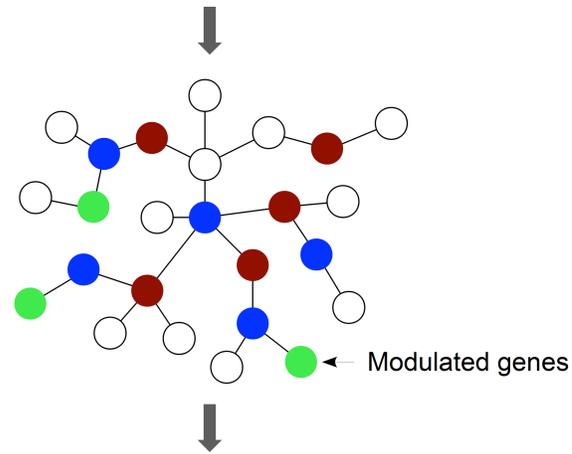
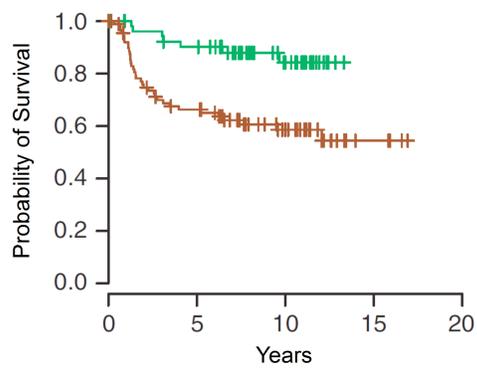